\documentclass[
]{jss}

\usepackage[utf8]{inputenc}

\providecommand{\tightlist}{%
  \setlength{\itemsep}{0pt}\setlength{\parskip}{0pt}}

\author{
Sahir Rai Bhatnagar*\\McGill University \And Maxime Turgeon*\\University of Manitoba \AND Jesse Islam\\McGill University \And James A. Hanley\\McGill University \And Olli Saarela\\University of Toronto
}
\title{\pkg{casebase}: An Alternative Framework For Survival Analysis and
Comparison of Event Rates}

\Plainauthor{Sahir Rai Bhatnagar*, Maxime Turgeon*, Jesse Islam, James A. Hanley, Olli Saarela}
\Plaintitle{casebase: An Alternative Framework For Survival Analysis and Comparison
of Event Rates}
\Shorttitle{\pkg{casebase}: An Alternative Framework For Survival Analysis}

\Abstract{
In epidemiological studies of time-to-event data, a quantity of interest
to the clinician and the patient is the risk of an event given a
covariate profile. However, methods relying on time matching or risk-set
sampling (including Cox regression) eliminate the baseline hazard from
the likelihood expression or the estimating function. The baseline
hazard then needs to be estimated separately using a non-parametric
approach. This leads to step-wise estimates of the cumulative incidence
that are difficult to interpret. Using case-base sampling, Hanley \&
Miettinen (2009) explained how the parametric hazard functions can be
estimated using logistic regression. Their approach naturally leads to
estimates of the cumulative incidence that are smooth-in-time.\\
In this paper, we present the \textbf{casebase} R package, a
comprehensive and flexible toolkit for parametric survival analysis. We
describe how the case-base framework can also be used in more complex
settings: competing risks, time-varying exposure, and variable
selection. Our package also includes an extensive array of visualization
tools to complement the analysis of time-to-event data. We illustrate
all these features through four different case studies.\\
*SRB and MT contributed equally to this work.
}

\Keywords{survival analysis, absolute risk, \proglang{R}, data visualization}
\Plainkeywords{survival analysis, absolute risk, R, data visualization}

\Address{
    Sahir Rai Bhatnagar*\\
    McGill University\\
    1020 Pine Avenue West Montreal, QC, Canada H3A 1A2\\
  E-mail: \email{sahir.bhatnagar@mail.mcgill.ca}\\
  URL: \url{http://sahirbhatnagar.com/}\\~\\
      Maxime Turgeon*\\
    University of Manitoba\\
    186 Dysart Road Winnipeg, MB, Canada R3T 2N2\\
  E-mail: \email{max.turgeon@umanitoba.ca}\\
  URL: \url{https://maxturgeon.ca/}\\~\\
      Jesse Islam\\
    McGill University\\
    1020 Pine Avenue West Montreal, QC, Canada H3A 1A2\\
  E-mail: \email{jesse.islam@mail.mcgill.ca}\\

      James A. Hanley\\
    McGill University\\
    1020 Pine Avenue West Montreal, QC, Canada H3A 1A2\\
  E-mail: \email{james.hanley@mcgill.ca}\\
  URL: \url{http://www.medicine.mcgill.ca/epidemiology/hanley/}\\~\\
      Olli Saarela\\
    University of Toronto\\
    Dalla Lana School of Public Health, 155 College Street, 6th floor,
Toronto, Ontario M5T 3M7, Canada\\
  E-mail: \email{olli.saarela@utoronto.ca}\\
  URL: \url{http://individual.utoronto.ca/osaarela/}\\~\\
  }

\usepackage{amsmath} \usepackage{amssymb} \usepackage{longtable} \usepackage{graphicx} \usepackage{tabularx} \usepackage{float} \usepackage{booktabs} \usepackage{makecell} \usepackage{tabu} \DeclareUnicodeCharacter{2500}{-}

\begin{document}

\hypertarget{introduction}{%
\section{Introduction}\label{introduction}}

Survival analysis and the comparison of event rates has been greatly
influenced over the last 50 years by the partial likelihood approach of
the Cox proportional hazard model \citep{cox1972regression}. This
approach provides a flexible way of assessing the influence of
covariates on the hazard function, without the need to specify a
parametric survival model. This flexibility comes at the cost of
decoupling the baseline hazard from the effect of the covariates. To
recover the whole survival curve---or the cumulative incidence function
(CIF)---we then need to separately estimate the baseline hazard
\citep{breslow1972discussion}. This in turn leads to stepwise estimates
of the survival function that can be difficult to interpret.

From the perspective of clinicians and their patients, the most relevant
quantity is often the 5- or 10-year risk of experiencing a certain event
given the patient's particular circumstances, and not the hazard ratio
between a treatment and control group. Therefore, to make sound clinical
decisions, it is important to accurately estimate the \emph{full} hazard
function, which can subsequently be used to estimate the cumulative
incidence function (CIF). Using a parametric estimator of the hazard
function leads to a smooth function of time; as a consequence, the CIF
and the survival function estimates also vary smoothly over time.

With the goal of fitting smooth-in-time hazard functions, Hanley \&
Miettinen \citeyearpar{hanley2009fitting} proposed a general framework
for estimating fully parametric hazard models via logistic regression.
Their main idea is simple: comparing person-moments when the event of
interest occurred with moments when patients were at risk. Their
approach handles censored data effortlessly and provides users familiar
with generalized linear models a natural way of fitting parametric
survival models. Moreover, their framework is very flexible: general
functions of time can be estimated (e.g.~using splines or general
additive models), and hence these models retain some of the flexibility
of Cox's partial likelihood approach. And since the unit of analysis is
a person moment, time-varying covariates can also easily be included in
this framework.

Logistic regression had already been used in the context of
discrete-time survival modeling \citep{cox1972regression}. But in the
context of continuous-time survival modeling, using the framework of
logistic regression opens the door to an extensive array of powerful
modeling tools. Indeed, lasso and elastic-net regression can be used to
select variables that are associated with the hazard function. Logistic
regression can also be replaced by multinomial regression to analyse
data with multiple competing events.

In this article, we present the \pkg{casebase} package
\citep{casebase-package} implemented in \proglang{R} \citep{r-core}, for
parametric survival analysis that combines the ideas of Hanley \&
Miettinen \citeyearpar{hanley2009fitting} into a simple interface. The
purpose of the \pkg{casebase} package is to provide practitioners with
an easy-to-use software tool to compute a patient's risk (or cumulative
incidence) of an event, conditional on a particular patient's covariate
profile. Our package retains the flexibility of case-base sampling and
the familiar interface of the \code{glm} function. It also provides
tools for variable selection and competing-risk analysis. In addition,
we provide extensive visualization tools.

In what follows, we first recall some theoretical details on case-base
sampling and its use for estimating parametric hazard functions. We then
give a short review of existing \proglang{R} packages that implement
comparable features as \pkg{casebase}. Next, we provide some details
about the implementation of case-base sampling in our package, and we
give a brief survey of its main functions. This is followed by four case
studies that illustrate the flexibility and capabilities of
\pkg{casebase}. We show how the same framework can be used for competing
risk analyses, penalized estimation, and for studies with time-dependent
exposures. Finally, we end the article with a discussion of the results
and of future directions.

\hypertarget{theory}{%
\section{Theoretical details}\label{theory}}

As discussed in Hanley \& Miettinen \citeyearpar{hanley2009fitting}, the
key idea behind case-base sampling is to sample from the study base a
finite amount of \emph{person moments}. These person moments are indexed
by both an individual in the study and a time point, and therefore each
person moment has a covariate profile, an exposure status and an outcome
status attached to it. We note that there is only a finite number of
person moments associated with the event of interest (what Hanley \&
Miettinen call the \emph{case series}). Case-base sampling refers to the
sampling from the base of a representative finite sample called the
\emph{base series}.

\hypertarget{likelihood-and-estimating-function}{%
\subsection{Likelihood and estimating
function}\label{likelihood-and-estimating-function}}

To describe the theoretical foundations of case-base sampling, we use
the framework of counting processes. In what follows, we abuse notation
slightly and omit any mention of \(\sigma\)-algebras. Instead, following
Aalen \emph{et al} \citeyearpar{aalen2008survival}, we use the
placeholder ``past'' to denote the past history of the corresponding
process. The reader interested in more details can refer to Saarela \&
Arjas \citeyearpar{saarela2015non} and Saarela
\citeyearpar{saarela2016case}. First, let \(N_{i}(t) \in \{0, 1\}\) be
counting processes corresponding to the event of interest for individual
\(i=1, \ldots,n\). For simplicity, we will consider Type I censoring due
to the end of follow-up at time \(\tau\) (the general case of
non-informative censoring is treated in Saarela
\citeyearpar{saarela2016case}). We assume a continuous time model, which
implies that the counting process jumps are less than or equal to one.
We are interested in modeling the hazard functions \(\lambda_{i}(t)\) of
the processes \(N_i(t)\), and which satisfy
\[\lambda_{i}(t) dt = E[dN_{i}(t)\mid \mathrm{past}].\]

Next, we model the base series sampling mechanism using non-homogeneous
Poisson processes \(R_i(t) \in \{0, 1, 2, \ldots\}\), with the
person-moments where \(dR_i(t) = 1\) constituting the base series. The
process \(Q_{i}(t) = R_i(t) + N_{i}(t)\) then counts both the case and
base series person-moments contributed by individual \(i\). This process
is typically defined by the user via its intensity function
\(\rho_i(t)\). The process \(Q_{i}(t)\) is characterized by
\(E[dQ_{i}(t)\mid\mathrm{past}] = \lambda_{i}(t)dt + \rho_i(t)dt\).

If the hazard function \(\lambda_{i}(t; \theta)\) is parametrized in
terms of \(\theta\), we could define an estimator \(\hat{\theta}\) by
maximization of the likelihood expression
\[L_0(\theta) = \prod_{i=1}^n \exp\left\{ -\int_0^{\min(t_i,\tau)} \lambda_i(t; \theta) dt \right\} \prod_{i=1}^{n} \prod_{t\in[0,\tau)} \lambda_{i}(t;\theta)^{dN_{i}(t)},\]
where \(\prod_{t\in[0,u)}\) represents a product integral from \(0\) to
\(u\), and where \(t_i\) is the event time for individual \(i\).
However, the integral over time makes the computation and maximization
of \(L_0(\theta)\) challenging.

Case-base sampling allows us to avoid this integral. By conditioning on
a sampled person-moment, we get individual likelihood contributions of
the form
\[P(dN_{i}(t) \mid dQ_{i}(t) = 1,\mathrm{past}) \stackrel{\theta}{\propto} \frac{\lambda_{i}(t; \theta)^{dN_{i}(t)}}{\rho_i(t) + \lambda_{i}(t;\theta)}.\]
Therefore, we can define an estimating function for \(\theta\) as
follows: \begin{equation}
L(\theta) = \prod_{i=1}^{n} \prod_{t\in[0,\tau)} \left(\frac{\lambda_{i}(t; \theta)^{dN_{i}(t)}}{\rho_i(t) + \lambda_{i}(t;\theta)}\right)^{dQ_i(t)}. \label{eq:lik-function}
\end{equation} When a logarithmic link function is used for modeling the
hazard function, the above expression is of a logistic regression form
with an offset term \(\log(1/\rho_i(t))\). Note that the sampling units
selected in the case-base sampling mechanism are person-moments, rather
than individuals, and the parameters to be estimated are hazards or
hazard ratios rather than odds or odds ratios. Generally, an individual
can contribute more than one person-moment, and thus the terms in the
product integral are not independent. Nonetheless, Saarela
\citeyearpar{saarela2016case} showed that the logarithm of this
estimating function has mean zero at the true value \(\theta=\theta_0\),
and that the resulting estimator \(\hat{\theta}\) is asymptotically
normally distributed.

In Hanley \& Miettinen \citeyearpar{hanley2009fitting}, the authors
suggest sampling the base series \emph{uniformly} from the study base.
In terms of Poisson processes, their sampling strategy corresponds
essentially to a time-homogeneous Poisson process with hazard equal to
\(\rho_i(t) = b/B\), where \(b\) is the number of sampled observations
in the base series, and \(B\) is the total population-time for the study
base (e.g.~the sum of all individual follow-up times). More complex
examples are also possible; see for example Saarela \& Arjas
\citeyearpar{saarela2015non}, where the intensity functions for the
sampling mechanism are proportional to the cardiovascular disease event
rate given by the Framingham score. Non-uniform sampling mechanisms can
increase the efficiency of the case-base estimators.

\hypertarget{common-parametric-models}{%
\subsection{Common parametric models}\label{common-parametric-models}}

Let \(g(t; X)\) be the linear predictor such that
\(\log(\lambda(t;X)) = g(t; X)\). Different functions of \(t\) lead to
different parametric hazard models. The simplest of these models is the
one-parameter exponential distribution which is obtained by taking the
hazard function to be constant over the range of \(t\):

\begin{equation}
\log(\lambda(t; X)) = \beta_0 + \beta_1 X. \label{eq:exp}
\end{equation}

In this model, the instantaneous failure rate is independent of
\(t\).\footnote{The conditional chance of failure in a time interval of specified length is the same regardless of how long the individual has been in the study. This is also known as the \textit{memoryless property} \citep{kalbfleisch2011statistical}.}

The Gompertz hazard model is given by including a linear term for time:

\begin{equation}
\log(\lambda(t; X)) = \beta_0 + \beta_1 t + \beta_2 X. \label{eq:gomp}
\end{equation}

Use of \(\log(t)\) yields the Weibull hazard which allows for a power
dependence of the hazard on time \citep{kalbfleisch2011statistical}:

\begin{equation}
\log(\lambda(t; X)) = \beta_0 + \beta_1 \log(t) + \beta_2 X. \label{eq:weibull}
\end{equation}

\hypertarget{competing-risk-analysis}{%
\subsection{Competing-risk analysis}\label{competing-risk-analysis}}

Case-base sampling can also be used in the context of competing-risk
analyses. Assuming there are \(J\) competing events, we can show that
each person-moment's contribution to the likelihood is of the form

\[\frac{\lambda_j(t)^{dN_j(t)}}{\rho(t) + \sum_{j=1}^J\lambda_j(t)},\]

where \(N_j(t)\) is the counting process associated with the event of
type \(j\) and \(\lambda_j(t)\) is the corresponding cause-specific
hazard function. As may be expected, this functional form is similar to
the terms appearing in the likelihood function for multinomial
regression.\footnote{Specifically, it corresponds to the following parametrization: \begin{align*} \log\left(\frac{P(Y=j \mid X)}{P(Y = J \mid X)}\right) = X^T\beta_j, \qquad j = 1,\ldots, J-1.\end{align*}}

\hypertarget{variable-selection}{%
\subsection{Variable selection}\label{variable-selection}}

To perform variable selection on the regression parameters
\(\theta \in \mathbb{R}^p\) of the hazard function, we can add a penalty
to the likelihood and optimise the following equation: \begin{equation}
\min _{\theta \in \mathbb{R}^{p}}\,\,-\ell\left(\theta\right)+\sum_{j=1}^p w_j P(\theta_j;\lambda,\alpha) \label{eq:penest}
\end{equation} where \(\ell\left(\theta\right) = \log L(\theta)\) is the
log of the likelihood function given in \eqref{eq:lik-function},
\(P(\theta_j;\lambda,\alpha)\) is a penalty term controlled by the
non-negative regularization parameters \(\lambda\) and \(\alpha\), and
\(w_j\) is the penalty factor for the \(j\)th covariate. These penalty
factors serve as a way of allowing parameters to be penalized
differently. For example, we could set the penalty factor for time to be
0 to ensure it is always included in the selected model.

\hypertarget{existing-packages}{%
\section{Existing packages}\label{existing-packages}}

Survival analysis is an important branch of applied statistics and
epidemiology. Accordingly, there is already a vast ecosystem of
\proglang{R} packages implementing different methodologies. In this
section, we describe how the functionalities of \pkg{casebase} compare
to these packages.

At the time of writing, a cursory examination of CRAN's task view on
survival analysis reveals that there are over 250 packages related to
survival analysis \citep{survTaskView}. For the purposes of this
article, we restricted our review to packages that implement at least
one of the following features: parametric modeling, non-proportional
hazard models, competing risk analysis, penalized estimation, and CIF
estimation. By searching for appropriate keywords in the
\code{DESCRIPTION} file of these packages, we found 60 relevant
packages. These 60 packages were then manually examined to determine
which ones are comparable to \pkg{casebase}. In particular, we excluded
packages that were focused on a different set of problems, such as
frailty and multi-state models. The remaining 14 packages appear in
Table \ref{tab:surv-pkgs}, along with some of the functionalities they
offer.

Parametric survival models are implemented in a handful of packages:
\pkg{CFC} \citeyearpar{mahani2015bayesian}, \pkg{flexsurv}
\citeyearpar{flexsurv}, \pkg{SmoothHazard} \citeyearpar{smoothHazard},
\pkg{rsptm2} \citeyearpar{clements_liu}, \pkg{mets}
\citeyearpar{scheike2014estimating}, and \pkg{survival}
\citeyearpar{survival-package}. The types of models they allow vary for
each package. For example, \pkg{SmoothHazard} is limited to Weibull
distributions \citeyearpar{smoothHazard}, whereas both \pkg{flexsurv}
and \pkg{survival} allow users to supply any distribution of their
choice. Also, \pkg{flexsurv}, \pkg{smoothhazard}, \pkg{mets} and
\pkg{rstpm2} also have the ability to model the effect of time using
splines, which allows flexible modeling of the hazard function.
Moreover, \pkg{flexsurv} has the ability to estimate both scale and
shape parameters for a variety of parametric families. As discussed
above, \pkg{casebase} can model any parametric family whose log-hazard
can be expressed as a linear combination of covariates (including time).
Therefore, our package allows the user to model the effect of time using
splines. Also, by including interaction terms between covariates and
time, it also allows users to fit (non-proportional) time-varying
coefficient models. However, we do not explicitly model any shape
parameter, unlike \pkg{flexsurv}.

Several packages implement penalized estimation for the Cox model:
\pkg{glmnet} \citeyearpar{regpathcox}, \pkg{glmpath}
\citeyearpar{park_hastie}, \pkg{penalized} \citeyearpar{l1penal},
\pkg{riskRegression} \citeyearpar{gerds_blanche}. Moreover, some
packages also include penalized estimation in the context of Cox models
with time-varying coefficients: elastic-net penalization with
\pkg{CoxRidge} \citeyearpar{perperoglou} and \pkg{rstpm2}
\citeyearpar{clements_liu}, while \pkg{survival}
\citeyearpar{survival-package} has an implementation of ridge-penalized
estimation. On the other hand, our package \pkg{casebase} provides
penalized estimation of the hazard function. To our knowledge,
\pkg{casebase} and \pkg{rsptm2} are the only packages to offer this
functionality.

Next, several \proglang{R} packages implement methodologies for
competing risk analysis; for a different perspective on this topic, see
Mahani \& Sharabiani \citeyearpar{mahani2015bayesian}. The package
\pkg{cmprsk} provides methods for cause-specific subdistribution
hazards, such as in the Fine-Gray model
\citeyearpar{fine1999proportional}. On the other hand, the package
\pkg{CFC} estimates cause-specific CIFs from unadjusted, non-parametric
survival functions. Our package \pkg{casebase} also provides
functionalities for competing risk analysis by estimating parametrically
the cause-specific hazards. From these quantities, we can then estimate
the cause-specific CIFs.

Finally, several packages include functions to estimate the CIF. The
corresponding methods generally fall into two categories: transformation
of the estimated hazard function, and semi-parametric estimation of the
baseline hazard. The first category broadly corresponds to parametric
survival models, where the full hazard is explicitly modeled. Using this
estimate, the survival function and the CIF can be obtained using their
functional relationships (see Equations \ref{eqn:surv} and \ref{eqn:CI}
below). Packages providing this functionality include \pkg{CFC},
\pkg{flexsurv}, \pkg{mets}, and \pkg{survival}. Our package
\pkg{casebase} also follows this approach for both single-event and
competing-risk analyses. The second category outlined above broadly
corresponds to semi-parametric models. These models do not model the
full hazard function, and therefore the baseline hazard needs to be
estimated separately in order to estimate the survival function. This is
achieved using semi-parametric estimators (e.g.~Breslow's estimator) or
parametric estimators (e.g.~spline functions). Packages that implement
this approach include \pkg{riskRegression}, \pkg{rstpm2}, and
\pkg{survival}. As mentioned in the introduction, a key distinguishing
factor between these two approaches is that the first category leads to
smooth estimates of the cumulative incidence function, whereas the
second category often produces estimates in the form of stepwise
functions. Providing smooth estimates of the CIF was one of the main
motivations for introducing case-base sampling in survival analysis.

\begin{table}[ht]
\centering
\resizebox{\textwidth}{!}{%
\begin{tabular}{ccccccccc}
\toprule
& \textbf{Competing} & \textbf{Allows} &  \textbf{Penalized}   &    &  & \textbf{Semi} & \textbf{Interval/Left} & \textbf{Absolute}  \\
\textbf{Package}        & \textbf{Risks}  & \textbf{Non PH} & \textbf{Regression} & \textbf{Splines} & \textbf{Parametric} & \textbf{Parametric} & \textbf{Censoring} & \textbf{Risk}           \\
\bottomrule
\textbf{casebase}        & \checkmark                        & \checkmark                         & \checkmark                     & \checkmark                & \checkmark                   &                          &                                  & \checkmark                                \\ \hline
\textbf{CFC}             & \checkmark                        & \checkmark                         &                       &                  & \checkmark                   &                          &                                  & \checkmark                                \\ \hline
\textbf{cmprsk}             & \checkmark                        &                           &                       &                  &                     &  \checkmark                        &                                  & \checkmark                                \\ \hline
\textbf{CoxRidge}        &                          & \checkmark                         & \checkmark                     &                  &                     & \checkmark                        &                                  &                                  \\ \hline
\textbf{crrp}            & \checkmark                        &                           & \checkmark                     &                  &                     & \checkmark                        &                                  &                                  \\ \hline
\textbf{fastcox}         &                          &                           & \checkmark                     &                  &                     & \checkmark                        &                                  &                                  \\ \hline
\textbf{flexrsurv}       &                          & \checkmark                         &                       & \checkmark                & \checkmark                   &                          &                                  & \checkmark               \\ \hline
\textbf{flexsurv}        & \checkmark                        & \checkmark                         &                       & \checkmark                & \checkmark                   &                          &                                  & \checkmark               \\ \hline
\textbf{glmnet}          &                          &                           & \checkmark                     &                  &                     & \checkmark                        &                                  &                                  \\ \hline
\textbf{glmpath}         &                          &                           & \checkmark                     &                  &                     & \checkmark                        &                                  &                                  \\ \hline
\textbf{mets}            & \checkmark                        &                           &                       & \checkmark                &                     & \checkmark                        &                                  & \checkmark                                \\ \hline
\textbf{penalized}       &                          &                           & \checkmark                     &                  &                     & \checkmark                        &                                  &                                  \\ \hline
\textbf{riskRegression}  & \checkmark                         &                           & \checkmark                     &                  &                     & \checkmark                        &                                  & \checkmark                                \\ \hline
\textbf{rstpm2}          &                          & \checkmark                         &                      & \checkmark                & \checkmark                   & \checkmark                        & \checkmark                                & \checkmark                         \\ \hline
\textbf{SmoothHazard}    &                          & \checkmark                         &                       & \checkmark                & \checkmark                   &                          & \checkmark                            &                                      \\ \hline
\textbf{survival}        & \checkmark                        & \checkmark                         &                       &                  & \checkmark                   & \checkmark                        & \checkmark                                & \checkmark                               \\
\bottomrule
\end{tabular}%
}
\caption{Comparison of various \proglang{R} packages for survival analysis based on several defining features. \textbf{Competing Risks}: whether or not an implementation for competing risks is present. \textbf{Allows Non PH}: permits models for non-proportional hazards. \textbf{Penalized Regression}: allows for a penalty term on the regression coefficients when estimating hazards (e.g. lasso or ridge). \textbf{Splines}: permits a flexible fit on time through the use of splines. \textbf{Parametric}: implementation for parametric models. \textbf{Semi-parametric}: implementation for semi-parametric models. \textbf{Interval/left censoring}: models for interval and left-censoring. If this is not selected, the package only handles right-censoring. \textbf{Absolute Risk}: computation for survival curves, cumulative incidence or cumulative hazard is readily available.}
\label{tab:surv-pkgs}
\end{table}

\hypertarget{implementation-details}{%
\section{Implementation details}\label{implementation-details}}

The functions in the \pkg{casebase} package can be divided into two
categories: 1) exploratory data analysis, in the form of population-time
plots; and 2) parametric modeling of the hazard function. We strove for
compatibility with both \code{data.frame}s and \code{data.table}s; this
can be seen in the coding choices we made and the unit tests we wrote.

\hypertarget{population-time-plots}{%
\subsection{Population-time plots}\label{population-time-plots}}

Population-time plots are a descriptive visualization of incidence
density, where aggregate person-time is represented by area and events
as points within the area. The case-base sampling approach described in
Section \ref{theory} can be visualized in the form of a population time
plot. These plots are extremely informative graphical displays of
survival data and should be one of the first steps in an exploratory
data analysis. The \code{popTime} function and \code{plot} method
facilitate this task:

\begin{enumerate}
\def\labelenumi{\arabic{enumi}.}
\tightlist
\item
  The \code{casebase::popTime} function takes as input the original
  dataset along with the column names corresponding to the timescale,
  the event status and an exposure group of interest (optional). This
  will create an object of class \code{popTime}.\\
\item
  The corresponding \code{plot} method for the object created in Step 1
  can be called to create the population time plot with several options
  for customizing the aesthetics.
\end{enumerate}

By splitting these tasks, we give flexibility to the user. While the
method call in Step 2 allows further customization by using the
\pkg{ggplot2} \citep{ggplot2} family of functions, users may choose the
graphics system of their choice to create population-time plots from the
object created in Step 1.

To illustrate these functions, we will use data from the European
Randomized Study of Prostate Cancer Screening (ERSPC)
\citep{schroder2009screening} which was extracted using the approach
described in Liu \emph{et al.} \citeyearpar{liu2014recovering}. This
dataset is available through the \pkg{casebase} package. It contains the
individual observations for 159,893 men from seven European countries,
who were between the ages of 55 and 69 years when recruited for the
trial.

We first create the necessary dataset for producing the population time
plot using the \code{popTime} function. In this example, we stratify the
plot by treatment group. The resulting object inherits from class
\code{popTime} and stores the exposure variable as an attribute:

\begin{CodeChunk}

\begin{CodeInput}
R> pt_object <- casebase::popTime(ERSPC, time = "Follow.Up.Time",
+                                event = "DeadOfPrCa", exposure = "ScrArm")
R> inherits(pt_object, "popTime")
\end{CodeInput}

\begin{CodeOutput}
#> [1] TRUE
\end{CodeOutput}

\begin{CodeInput}
R> attr(pt_object, "exposure")
\end{CodeInput}

\begin{CodeOutput}
#> [1] "ScrArm"
\end{CodeOutput}
\end{CodeChunk}

We then pass this object to the corresponding \code{plot} method:

\begin{CodeChunk}

\begin{CodeInput}
R> plot(pt_object, add.base.series = TRUE)
\end{CodeInput}
\end{CodeChunk}

\begin{CodeChunk}
\begin{figure}[ht]

{\centering \includegraphics[width=\textwidth,keepaspectratio=true]{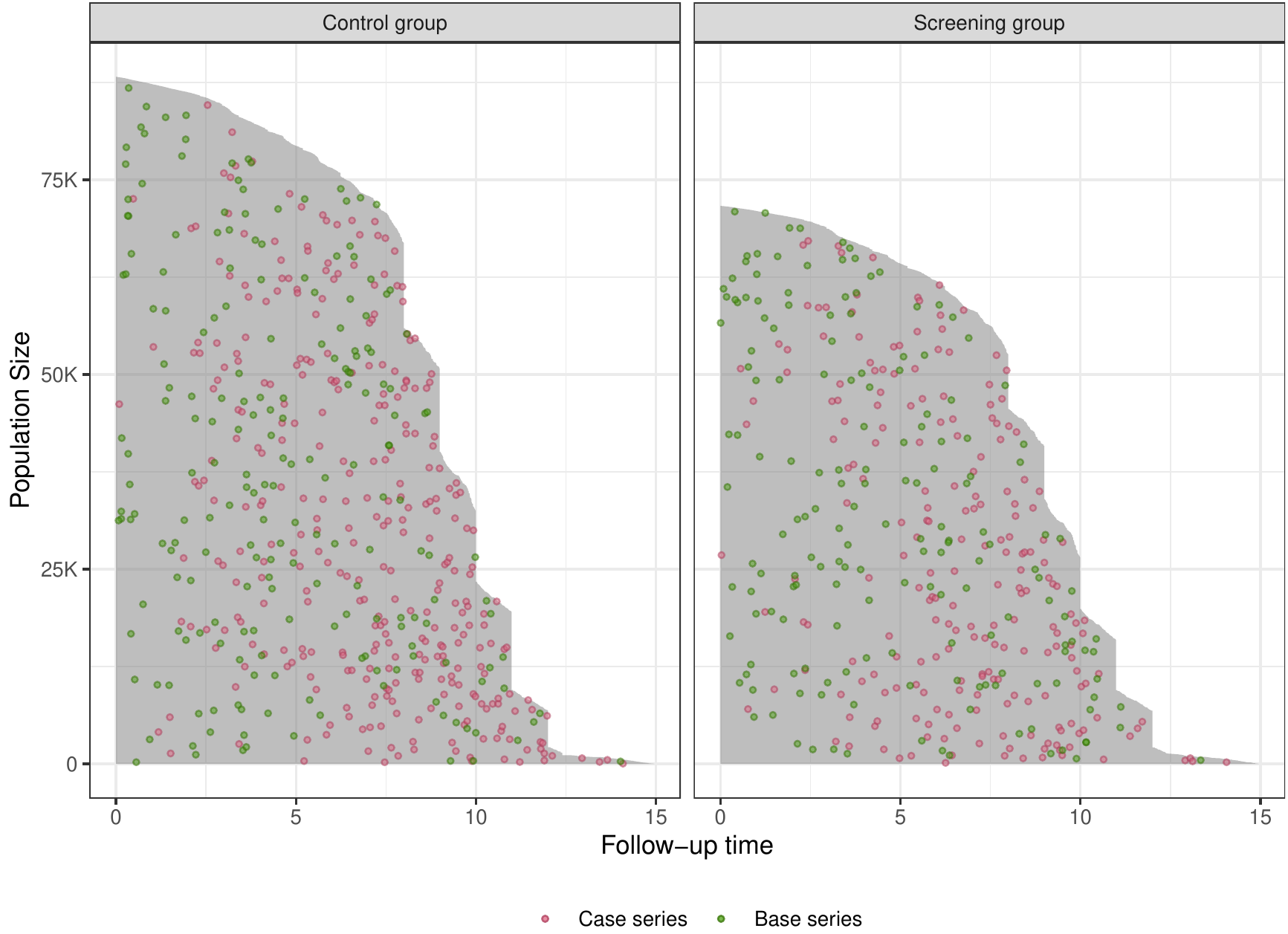}

}

\caption[Population time plots for both treatment arms in the ERSPC dataset]{Population time plots for both treatment arms in the ERSPC dataset. The gray area can be thought of as N=88,232 (control group) and N=71,661 (screening group) rows of infinitely thin rectangles (person-moments). More events are observed at later follow-up times, motivating the use of non-constant hazard models.}\label{fig:plot-stratified-erspc-data}
\end{figure}
\end{CodeChunk}

Figure \ref{fig:plot-stratified-erspc-data} is built sequentially by
first adding a layer for the area representing the population time in
gray, with subjects having the least amount of observation time plotted
at the top of the y-axis. We immediately notice a distinctive
\emph{stepwise shape} in the population time area. This is due to the
randomization of the Finnish cohorts which were carried out on January 1
of each of year from 1996 to 1999. Coupled with the uniform December 31
2006 censoring date, this led to large numbers of men with exactly 11,
10, 9 or 8 years of follow-up. Tracked backwards in time (i.e.~from
right to left), the population-time plot shows the recruitment pattern
from its beginning in 1991, and the January 1 entries in successive
years. Tracked forwards in time (i.e.~from left to right), the plot for
the first three years shows attrition due entirely to death (mainly from
other causes). Since the Swedish and Belgian centres were the last to
complete recruitment in December 2003, the minimum potential follow-up
is three years. Tracked further forwards in time (i.e.~after year 3) the
attrition is a combination of deaths and staggered entries. As we can
see, population-time plots summarise a wealth of information about the
study into a simple graph.

Next, layers for the case series and base series are added. The y-axis
location of each case moment is sampled at random vertically on the plot
to avoid having all points along the upper edge of the gray area. By
randomly distributing the cases, we can get a sense of the incidence
density. In Figure \ref{fig:plot-stratified-erspc-data}, we see that
more events are observed at later follow-up times. Therefore, a constant
hazard model would not be appropriate in this instance as it would
overestimate the cumulative incidence earlier on in time, and
underestimate it later on. Finally, the base series is sampled
horizontally with sampling weight proportional to their follow-up time.
The reader should refer to the package vignettes for more examples and a
detailed description of how to modify the aesthetics of a
population-time plot.

\hypertarget{parametric-modeling}{%
\subsection{Parametric modeling}\label{parametric-modeling}}

The parametric modeling step was separated into three parts:

\begin{enumerate}
\def\labelenumi{\arabic{enumi}.}
\tightlist
\item
  case-base sampling;
\item
  estimation of the smooth hazard function;
\item
  estimation of the CIF.
\end{enumerate}

By separating the sampling and estimation functions, we allow the
possibility of users implementing more complex sampling scheme (as
described in Saarela \citeyearpar{saarela2016case}), or more complex
study designs (e.g.~time-varying exposure).

The sampling scheme selected for \code{sampleCaseBase} was described in
Hanley \& Miettinen \citeyearpar{hanley2009fitting}: we first sample
along the ``person'' axis, proportional to each individual's total
follow-up time, and then we sample a moment uniformly over their
follow-up time. This sampling scheme is equivalent to the following
picture: imagine representing the total follow-up time of all
individuals in the study along a single dimension, where the follow-up
time of the next individual would start exactly when the follow-up time
of the previous individual ends. Then the base series could be sampled
uniformly from this one-dimensional representation of the overall
follow-up time. In any case, the output is a dataset of the same class
as the input, where each row corresponds to a person-moment. The
covariate profile for each such person-moment is retained, and an offset
term is added to the dataset. This output could then be used to fit a
smooth hazard function, or for visualization of the base series.

Next, the fitting function \code{fitSmoothHazard} starts by looking at
the class of the dataset: if it was generated from
\code{sampleCaseBase}, it automatically inherited the class
\code{cbData}. If the dataset supplied to \code{fitSmoothHazard} does
not inherit from \code{cbData}, then the fitting function starts by
calling \code{sampleCaseBase} to generate the base series. In other
words, users can bypass \code{sampleCaseBase} altogether and only worry
about the fitting function \code{fitSmoothHazard}.

The fitting function retains the familiar formula interface of
\code{glm}. The left-hand side of the formula should be the name of the
column corresponding to the event type. The right-hand side can be any
combination of the covariates, along with an explicit functional form
for the time variable. Note that non-proportional hazard models can be
achieved at this stage by adding an interaction term involving time
(cf.~Case Study 4 below). The offset term does not need to be specified
by the user, as it is automatically added to the formula before calling
\code{glm}.

To fit the hazard function, we provide several approaches that are
available via the \code{family} parameter. These approaches are:

\begin{itemize}
\tightlist
\item
  \code{glm}: This is the familiar logistic regression.
\item
  \code{glmnet}: This option allows for variable selection using the
  elastic-net \citep{zou2005regularization} penalty (cf.~Case Study 3).
  This functionality is provided through the \pkg{glmnet} package
  \citep{friedman2010jss}.
\item
  \code{gam}: This option provides support for \emph{Generalized
  Additive Models} via the \pkg{mgcv} package
  \citep{hastie1987generalized}.
\item
  \code{gbm}: This option provides support for \emph{Gradient Boosted
  Trees} via the \pkg{gbm} package. This feature is still experimental.
\end{itemize}

In the case of multiple competing events, the hazard is fitted via
multinomial regression as performed by the \pkg{VGAM} package. We
selected this package for its ability to fit multinomial regression
models with an offset.

Once a model-fit object has been returned by \code{fitSmoothHazard}, all
the familiar summary and diagnostic functions are available:
\code{print}, \code{summary}, \code{predict}, \code{plot}, etc. Our
package provides one more functionality: it computes risk functions from
the model fit. For the case of a single event, it uses the familiar
identity \begin{equation}\label{eqn:surv}
S(t) = \exp\left(-\int_0^t \lambda(u;X) du\right).
\end{equation} The integral is computed using either the numerical or
Monte-Carlo integration. The risk function (or cumulative incidence
function) is then defined as \begin{equation}\label{eqn:CI}
CI(t) = 1 - S(t).
\end{equation}

For the case of a competing-event analysis, the event-specific risk is
computed using the following procedure: first, we compute the overall
survival function (i.e.~for all event types):

\[ S(t) = \exp\left(-\int_0^t \lambda(u;X) du\right),\qquad \lambda(t;X) = \sum_{j=1}^J \lambda_j(t;X).\]
From this, we can derive the event-specific subdensities:

\[ f_j(t) = \lambda_j(t)S(t).\]

Finally, by integrating these subdensities, we obtain the event-specific
cumulative incidence functions:

\[ CI_j(t) = \int_0^t f_j(u)du.\] Again, he integrals are computed using
either numerical integration (via the trapezoidal rule) or Monte Carlo
integration. This option is controlled by the argument \code{method} of
the \code{absoluteRisk} function.

In the following sections, we illustrate these functionalities in the
context of four case studies.

\hypertarget{case-study-1european-randomized-study-of-prostate-cancer-screening}{%
\section{Case study 1---European Randomized Study of Prostate Cancer
Screening}\label{case-study-1european-randomized-study-of-prostate-cancer-screening}}

For our first case study, we return to the ERSPC study and investigate
the differences in risk between the control and screening arms.

\hypertarget{different-functional-forms-of-time}{%
\subsection{Different functional forms of
time}\label{different-functional-forms-of-time}}

We fit four models that differ in which functional form of time is used:
1) excluded from the linear predictor as seen in \eqref{eq:exp}, 2)
linear function as seen in \eqref{eq:gomp}, 3) log function as seen in
\eqref{eq:weibull}, and 4) a smooth function using cubic B-splines. The
models are fit using \code{fitSmoothHazard} with the familiar formula
interface:

\begin{CodeChunk}

\begin{CodeInput}
R> fmla <- list(exponential = as.formula(DeadOfPrCa ~ ScrArm),
+              gompertz = as.formula(DeadOfPrCa ~ Follow.Up.Time + ScrArm),
+              weibull = as.formula(DeadOfPrCa ~ log(Follow.Up.Time) + ScrArm),
+              splines = as.formula(DeadOfPrCa ~ bs(Follow.Up.Time) + ScrArm))
R>
R> fits <- lapply(fmla, function(i) {
+   fitSmoothHazard(i, data = ERSPC, ratio = 100)
+ })
\end{CodeInput}
\end{CodeChunk}

The output object from \code{fitSmoothHazard} inherits from the
\code{singleEventCB} and \code{glm} classes. As such, we can directly
use the \code{summary} generic:

\begin{CodeChunk}

\begin{CodeInput}
R> summary(fits[["splines"]])
\end{CodeInput}
\end{CodeChunk}

\begin{CodeChunk}

\begin{CodeOutput}
#>
#> Coefficients:
#>                       Estimate Std. Error z value Pr(>|z|)
#> (Intercept)           -10.2905     0.3185  -32.31  < 2e-16
#> bs(Follow.Up.Time)1     4.3587     0.8008    5.44  5.2e-08
#> bs(Follow.Up.Time)2     2.0890     0.4721    4.42  9.7e-06
#> bs(Follow.Up.Time)3     4.6266     0.6910    6.70  2.1e-11
#> ScrArmScreening group  -0.2354     0.0886   -2.66   0.0079
#>
#> (Dispersion parameter for binomial family taken to be 1)
#>
#>     Null deviance: 6059.0  on 54539  degrees of freedom
#> Residual deviance: 5785.3  on 54535  degrees of freedom
#> AIC: 5795
#>
#> Number of Fisher Scoring iterations: 9
\end{CodeOutput}
\end{CodeChunk}

Next, the \code{absoluteRisk} function takes as input the
\code{fitSmoothHazard} object and returns a matrix where each column
corresponds to the covariate profiles specified in the \code{newdata}
argument, and each row corresponds to a specified time point:

\begin{CodeChunk}

\begin{CodeInput}
R> new_data <- data.frame(ScrArm = c("Control group", "Screening group"))
R> new_time <- seq(0,14,0.1)
R>
R> risks <- lapply(fits, function(i) {
+   absoluteRisk(object = i, time = new_time, newdata = new_data, method = "mont")
+ })
\end{CodeInput}
\end{CodeChunk}

In Figure \ref{fig:erspc-cox-cif}, we overlay the estimated CIFs from
\pkg{casebase} on the Cox model CIF. The CIF estimates for the
exponential model in panel (1) overestimate the cumulative incidence
earlier on in time, and underestimate it later on. Based on our earlier
discussion of the population-time plot, this poor fit for the
exponential hazard was expected. We notice a better fit with increasing
complexity of our model for time in Figure \ref{fig:erspc-cox-cif}
(panels 2--4). As noted above, the usual asymptotic results hold for
likelihood ratio tests built using case-base sampling models. Therefore,
we can easily test the null hypothesis that the exponential model is
just as good as the larger (in terms of number of parameters) spline
model:

\begin{CodeChunk}

\begin{CodeOutput}
#> Analysis of Deviance Table
#>
#> Model 1: DeadOfPrCa ~ ScrArm + offset(offset)
#> Model 2: DeadOfPrCa ~ bs(Follow.Up.Time) + ScrArm + offset(offset)
#>   Resid. Df Resid. Dev Df Deviance Pr(>Chi)
#> 1     54538       6052
#> 2     54535       5785  3      267   <2e-16 ***
#> ---
#> Signif. codes:  0 '***' 0.001 '**' 0.01 '*' 0.05 '.' 0.1 ' ' 1
\end{CodeOutput}
\end{CodeChunk}

The null hypothesis is rejected in favor of the spline model. Similarly,
the AIC provides further evidence that the flexible function of time
provides the best fit:

\begin{CodeChunk}

\begin{CodeOutput}
#>     Exp. Gompertz  Weibull  Splines
#>     6056     5821     5807     5795
\end{CodeOutput}
\end{CodeChunk}

\begin{CodeChunk}
\begin{figure}[ht]
\includegraphics[width=\textwidth,keepaspectratio=true]{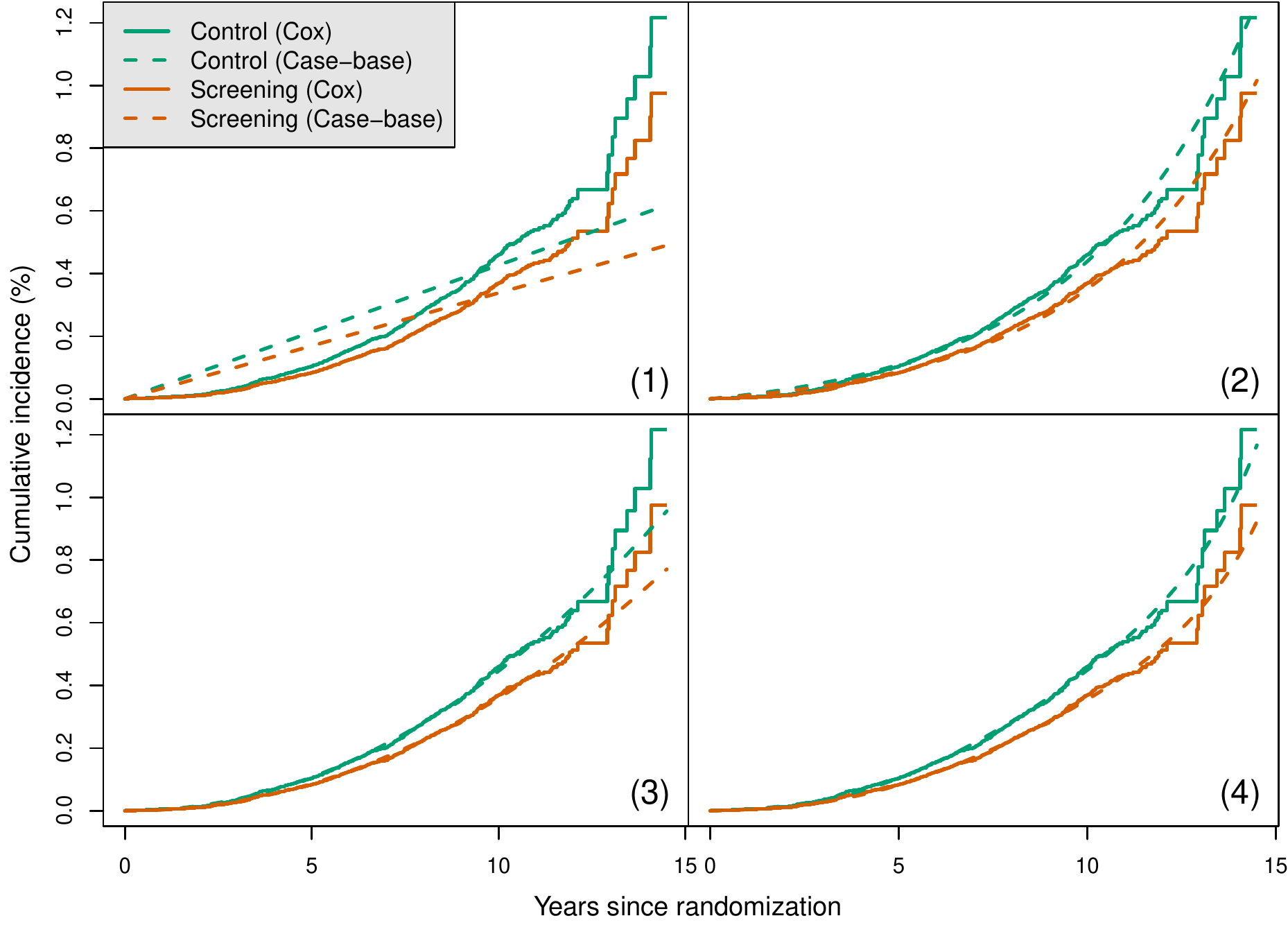} \caption{CIFs for control and screening groups in the ERSPC data. In each of the panels, we plot the CIF from the Cox model using \code{survival::survfit} (solid line) and the CIF from the case-base sampling scheme (dashed line) with different functional forms of time. (1) The time variable is excluded (exponential). (2) Linear function of time (Gompertz). (3) The natural logarithm (Weibull). (4) Cubic B-spline expansion of time.}\label{fig:erspc-cox-cif}
\end{figure}
\end{CodeChunk}

In Table \ref{tab:print-erspc-estimates}, we present a side-by-side
comparison of the hazard ratios and confidence intervals estimated from
\code{fitSmoothHazard} and the corresponding parametric model using
\code{survival::survreg}, as well as the Cox model estimate. The hazard
ratio estimates and confidence intervals are similar across all four
models. This reinforces the idea that, under proportional hazards, we do
not need to model the full hazard to obtain reliable estimates of the
hazard ratio. Nevertheless, Figure \ref{fig:erspc-cox-cif} shows that
different parametric models can still give rise to qualitatively
different estimates for the CIF.

\begin{CodeChunk}
\begin{table}

\caption{\label{tab:print-erspc-estimates}Comparison of estimated hazard ratios and 95\% confidence intervals for ERSPC data.}
\centering
\begin{tabular}[t]{>{}lcc}
\toprule
Model & casebase::fitSmoothHazard & survival::survreg\\
\midrule
\textbf{Exponential} & 0.79 (0.67, 0.94) & 0.81 (0.68, 0.96)\\
\cmidrule{1-3}
\textbf{Gompertz} & 0.80 (0.67, 0.95) & 0.80 (0.67, 0.95)\\
\cmidrule{1-3}
\textbf{Weibull} & 0.81 (0.68, 0.96) & 0.80 (0.65, 0.96)\\
\cmidrule{1-3}
\textbf{Splines} & 0.79 (0.66, 0.94) & --\\
\bottomrule
\multicolumn{3}{l}{\rule{0pt}{1em}Cox model estimate: HR (95\% CI) = 0.80 (0.67, 0.95)}\\
\end{tabular}
\end{table}

\end{CodeChunk}

\hypertarget{time-dependent-hazard-ratios}{%
\subsection{Time-dependent hazard
ratios}\label{time-dependent-hazard-ratios}}

Previous re-analyses of these data suggested that the overall screening
attributed reduction in death due to prostate cancer of 20\% was an
underestimate \citep{hanley2010mortality}. The estimated 20\% (from a
proportional hazards model) did not account for the delay between
screening and the time the effect is expected to be observed. As a
result, the null effects in years 1--7 masked the substantial reductions
that began to appear from year 8 onwards. This motivates the use of a
time-dependent hazard ratio which can easily be fit with the
\pkg{casebase} package by including an interaction term with time in the
model:

\begin{CodeChunk}

\begin{CodeInput}
R> fit_inter <- fitSmoothHazard(DeadOfPrCa ~ bs(Follow.Up.Time) * ScrArm,
+                              data = ERSPC)
\end{CodeInput}
\end{CodeChunk}

In Figure \ref{fig:interaction-ERSPC}, we have the estimated hazard
ratio and 95\% confidence interval for screening vs.~control group as a
function of time using the \code{plot} method for objects of class
\code{singleEventCB}:

\begin{CodeChunk}

\begin{CodeInput}
R> plot(fit_inter, type = "hr", newdata = new_data,
+      var = "ScrArm", xvar = "Follow.Up.Time", ci = TRUE)
\end{CodeInput}
\end{CodeChunk}

The plot shows that the cures attributable to the screening only begin
to become statistically apparent by year 7 and later. The 25-60\%
reductions seen in years 8-12 of the study suggests a much higher
reduction in prostate cancer due to screening than the single overall
20\% reported in the original article.

\begin{CodeChunk}
\begin{figure}[ht]

{\centering \includegraphics[width=\textwidth,keepaspectratio=true]{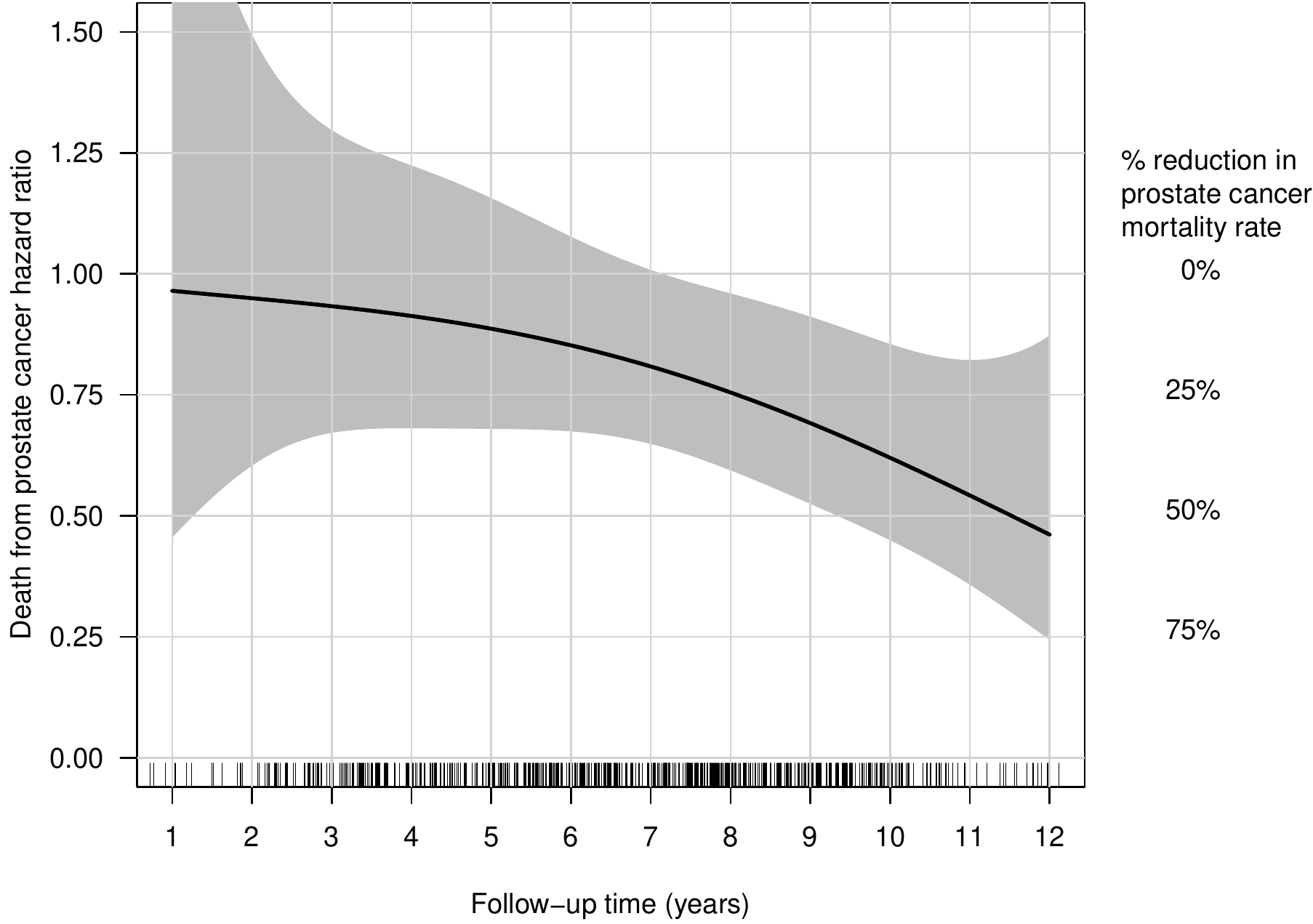}

}

\caption[Estimated hazard ratio and 95\% confidence interval for screening vs]{Estimated hazard ratio and 95\% confidence interval for screening vs. control group as a function of time in the ERSPC dataset. Hazard ratios are estimated from fitting a parametric hazard model as a function of the interaction between a cubic B-spline basis of follow-up time and treatment arm. 95\% confidence intervals are calculated using the delta method. The plot shows that the effect of screening only begins to become statistically apparent by year 7. The 25-60\% reductions seen in years 8-12 of the study suggests a much higher reduction in prostate cancer due to screening than the single overall 20\% reported in the original article.}\label{fig:interaction-ERSPC}
\end{figure}
\end{CodeChunk}

With this first case study, we explored how \pkg{casebase} allows us to
fit different parametric survival models with possible time-varying
effects, and how we can compare the fit of each model with tools from
GLMs.

\hypertarget{case-study-2bone-marrow-transplant}{%
\section{Case study 2---Bone-marrow
transplant}\label{case-study-2bone-marrow-transplant}}

In the next case study, we show how case-base sampling can be used in
the context of a competing-risk analysis. For illustrative purposes, we
use the same data that was used in Scrucca \emph{et al}
\citeyearpar{scrucca2010regression}. The data was downloaded from the
first author's website, and it is now available as part of the
\pkg{casebase} package.

The data contains information on 177 patients who received a stem-cell
transplant for acute leukemia. The event of interest is relapse, but
other competing causes (e.g.~death, progression, graft failure,
graft-versus-host disease) were also recorded. Several covariates were
captured at baseline: sex, disease type (acute lymphoblastic or
myeloblastic leukemia, abbreviated as ALL and AML, respectively),
disease phase at transplant (Relapse, CR1, CR2, CR3), source of stem
cells (bone marrow and peripheral blood, coded as BM+PB, or only
peripheral blood, coded as PB), and age.

First, we look at a population-time plot to visualize the incidence
density of both relapse and the competing events. In Figure
\ref{fig:compPop}, failure times are highlighted on the plot using red
dots for the event of interest (panel A) and blue dots for competing
events (panel B). In both panels, we see evidence of a non-constant
hazard function: the density of points is larger at the beginning of
follow-up than at the end.

\begin{CodeChunk}
\begin{figure}[ht]

{\centering \includegraphics[width=\textwidth,keepaspectratio=true]{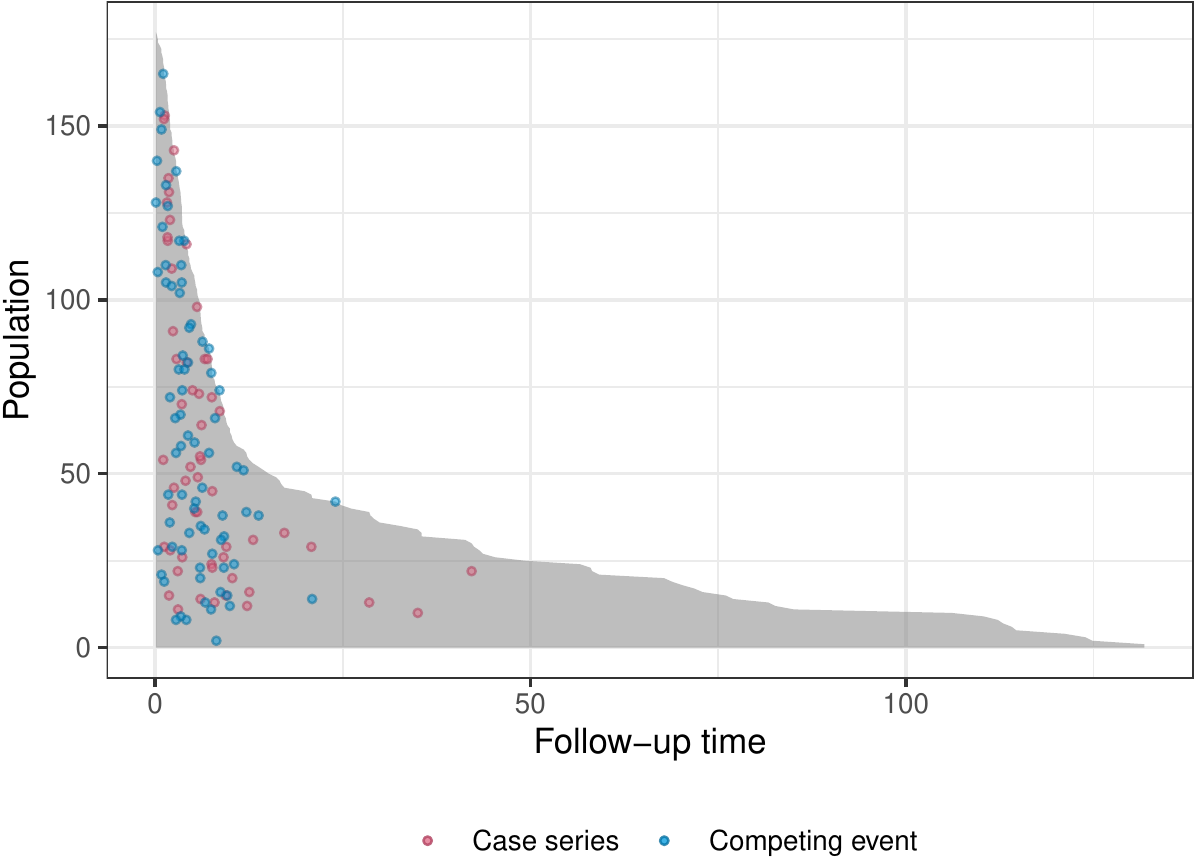}

}

\caption[Population-time plot for the stem-cell transplant study with both relapse and competing events]{Population-time plot for the stem-cell transplant study with both relapse and competing events.}\label{fig:compPop}
\end{figure}
\end{CodeChunk}

Our main objective is to compute the cumulative incidence of relapse for
a given set of covariates. We start by fitting a smooth hazard to the
data using a linear term for time:

We want to compare our hazard ratio estimates to that obtained from a
Cox regression. To obtain these estimates, we need to treat the
competing event as censoring:

\begin{CodeChunk}

\begin{CodeInput}
R> library(survival)
R> # Treat competing event as censoring
R> model_cox <- coxph(Surv(ftime, Status == 1) ~ Sex + D + Phase + Source + Age,
+                    data = bmtcrr)
\end{CodeInput}
\end{CodeChunk}

\begin{CodeChunk}
\begin{table}

\caption{\label{tab:bmtcrr-cis}Estimates and confidence intervals for the hazard ratios for each coefficient. Both estimates from case-base sampling and Cox regression are presented.}
\centering
\begin{tabular}[t]{lrlrl}
\toprule
\multicolumn{1}{c}{ } & \multicolumn{2}{c}{Case-Base} & \multicolumn{2}{c}{Cox} \\
\cmidrule(l{3pt}r{3pt}){2-3} \cmidrule(l{3pt}r{3pt}){4-5}
Covariates & HR & 95\% CI & HR & 95\% CI\\
\midrule
Sex & 0.74 & (0.43, 1.3) & 0.68 & (0.39, 1.19)\\
Disease & 0.52 & (0.29, 0.94) & 0.52 & (0.29, 0.93)\\
Phase (CR2 vs. CR1) & 1.20 & (0.48, 3) & 1.21 & (0.48, 3.02)\\
Phase (CR3 vs. CR1) & 1.59 & (0.41, 6.17) & 1.67 & (0.43, 6.5)\\
Phase (Relapse vs. CR1) & 4.09 & (1.9, 8.81) & 4.55 & (2.09, 9.9)\\
\addlinespace
Source & 1.70 & (0.55, 5.23) & 1.46 & (0.47, 4.54)\\
Age & 0.99 & (0.97, 1.02) & 0.99 & (0.97, 1.02)\\
\bottomrule
\end{tabular}
\end{table}

\end{CodeChunk}

From the fit object, we can extract both the hazard ratios and their
corresponding confidence intervals. These quantities appear in Table
\ref{tab:bmtcrr-cis}. As we can see, the only significant hazard ratio
identified by case-base sampling is the one associated with the phase of
the disease at transplant. More precisely, being in relapse at
transplant is associated with a hazard ratio of 3.89 when compared to
CR1.

Given the estimate of the hazard function obtained using case-base
sampling, we can compute the absolute risk curve for a fixed covariate
profile. We perform this computation for a 35 year old woman who
received a stem-cell transplant from peripheral blood at relapse. We
compare the absolute risk curve for such a woman with ALL with that for
a similar woman with AML.

Next, we compare our estimates to that obtained from a corresponding
Fine-Gray model \citeyearpar{fine1999proportional}. The Fine-Gray model
is a semiparametric model for the cause-specific \emph{subdistribution
hazard}, i.e.~the function \(d_k(t)\) such that
\[CI_k(t) =1 - \exp\left( - \int_0^t d_k(u) \textrm{d}u \right),\] where
\(CI_k(t)\) is the cause-specific cumulative incidence. The Fine-Gray
model allows to directly assess the effect of a covariate on the
subdistribution hazard, as opposed to the cause-specific hazard. For the
computation, we use the \pkg{timereg} package \citep{timereg}:

\begin{CodeChunk}

\begin{CodeInput}
R> library(timereg)
R> model_fg <- comp.risk(Event(ftime, Status) ~ const(Sex) + const(D) +
+                         const(Phase) + const(Source) + const(Age),
+                       data = bmtcrr, cause = 1, model = "fg")
R>
R> # Estimate absolute risk curve
R> risk_fg <- predict(model_fg, newdata, times = time_points)
\end{CodeInput}
\end{CodeChunk}

\begin{CodeChunk}
\begin{figure}[ht]

{\centering \includegraphics[width=\textwidth,keepaspectratio=true]{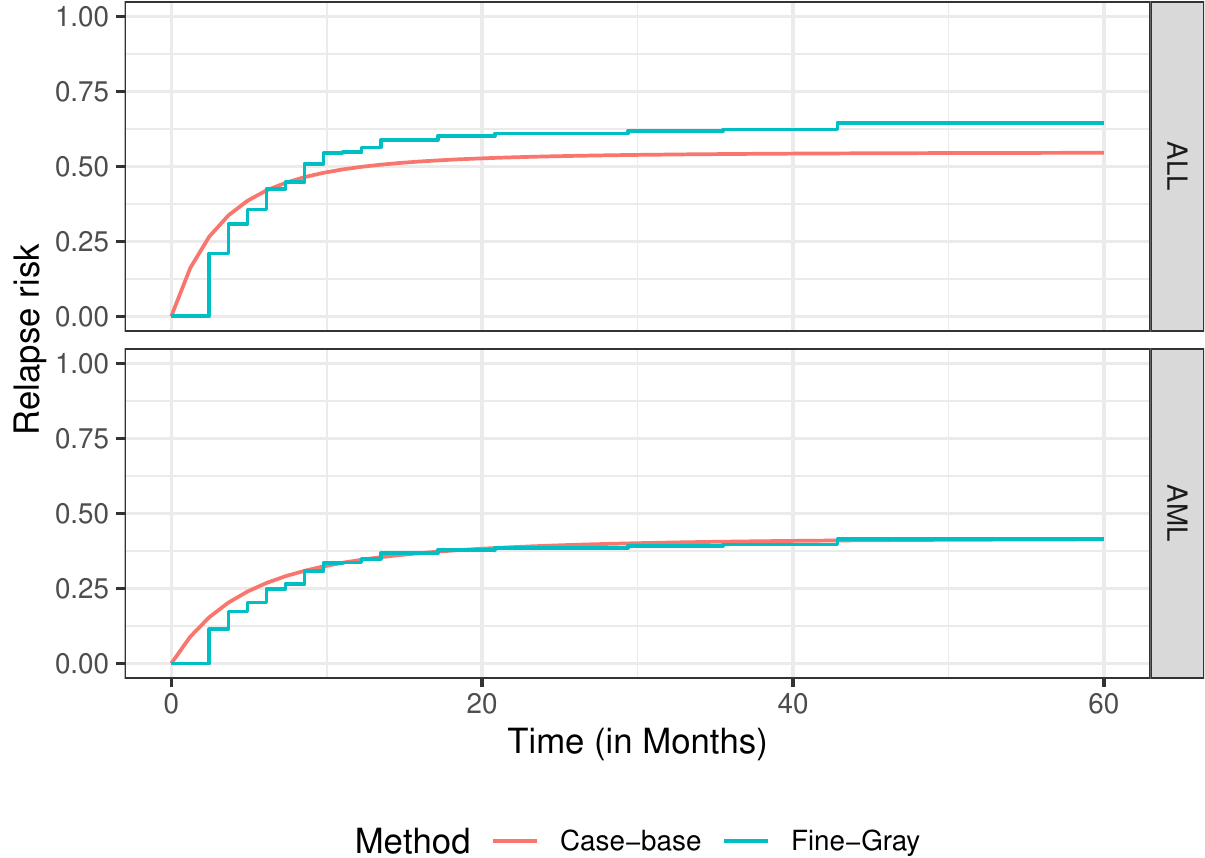}

}

\caption{\label{fig:compAbsrisk} Absolute risk curve for a fixed covariate profile and the two disease groups. The estimate obtained from case-base sampling is compared to the Kaplan-Meier estimate.}\label{fig:bmtcrr-risk}
\end{figure}
\end{CodeChunk}

Figure \ref{fig:compAbsrisk} shows the absolute risk curves for both
case-base sampling and the Fine-Gray model. As we can see, the two
approaches agree quite well for AML; however, there seems to be a
difference of about 5\% between the two curves for ALL. This difference
does not appear to be significant: the curve from case-base sampling is
contained within a 95\% confidence band around the Fine-Gray absolute
risk curve (figure not shown).

\hypertarget{case-study-3support-data}{%
\section{Case study 3---SUPPORT Data}\label{case-study-3support-data}}

In the first two case studies, we described the basic functionalities of
the \pkg{casebase} package: creating population-time plots, fitting
parametric models for hazard functions, and estimating the corresponding
cumulative incidence curves. For the third case study, we show how
\pkg{casebase} can also be used for variable selection through
regularized estimation of the hazard function as given by
\eqref{eq:penest}.

To illustrate this functionality, we use the dataset from the Study to
Understand Prognoses Preferences Outcomes and Risks of Treatment
(SUPPORT) \citep{knaus1995support}. The SUPPORT dataset tracks death in
five American hospitals within individuals who are considered seriously
ill.~The original data is available online from the Department of
Biostatistics at Vanderbilt University \citep{harrell_2020}. The cleaned
and imputed data consists of 9104 observations and 30 variables, and it
is available as part of the \pkg{casebase} package. In the comparisons
below, all covariates except \code{sps} and \code{aps} were used. These
two variables correspond to scores for predicting the outcome that were
developed as part of the original study. For more information about this
dataset, the reader is encouraged to look at the documentation in our
package.

For our penalized case-base model, we opt for the natural log of time
which corresponds to a Weibull distribution. For fitting the penalized
hazard, we use \code{fitSmoothHazard.fit}, which is a matrix interface
to the \code{fitSmoothHazard} function. We supply both a matrix \code{y}
containing the time and event variables, and a matrix \code{x}
containing all other covariates. We apply the lasso penalty by setting
\code{alpha = 1} and assign a \code{penalty.factor} (\(w_j\);
cf.~Equation \ref{eq:penest}) of 0 to the time variable to ensure it is
in the selected model. We compare our approach to both Cox regression,
and lasso penalized Cox regression (fitted via the \pkg{glmnet}
package).

To compare the performance of our models, we split the data into 95\%
training and 5\% test sets. To assess both discrimination and
calibration, we use a time-dependent version of the classical Brier
score that is adjusted for censoring \citep{graf1999ass}. The Brier
score can be used with both parametric and semi-parametric models. We
use the \pkg{riskRegression} package to compute these scores for all
models.

\begin{CodeChunk}

\begin{CodeInput}
R> # Create matrices for inputs
R> x <- model.matrix(death ~ . - d.time - aps - sps,
+                   data = train)[, -c(1)] # Remove intercept
R> y <- data.matrix(subset(train, select = c(d.time, death)))
R>
R> # Regularized logistic regression to estimate hazard
R> pen_cb <- casebase::fitSmoothHazard.fit(x, y,
+   family = "glmnet",
+   time = "d.time", event = "death",
+   formula_time = ~ log(d.time), alpha = 1,
+   ratio = 10, standardize = TRUE,
+   penalty.factor = c(0, rep(1, ncol(x)))
+ )
\end{CodeInput}
\end{CodeChunk}

In Figure \ref{fig:cs3lolliPlot}, we show the coefficient estimates for
covariates that we selected by both penalized Cox and penalized
case-base. We note that both penalized approaches produce similar
results. We can also clearly see the shrinkage effect owing to the
\(\ell_1\) penalty.

\begin{CodeChunk}
\begin{figure}[ht]

{\centering \includegraphics[width=\textwidth,keepaspectratio=true]{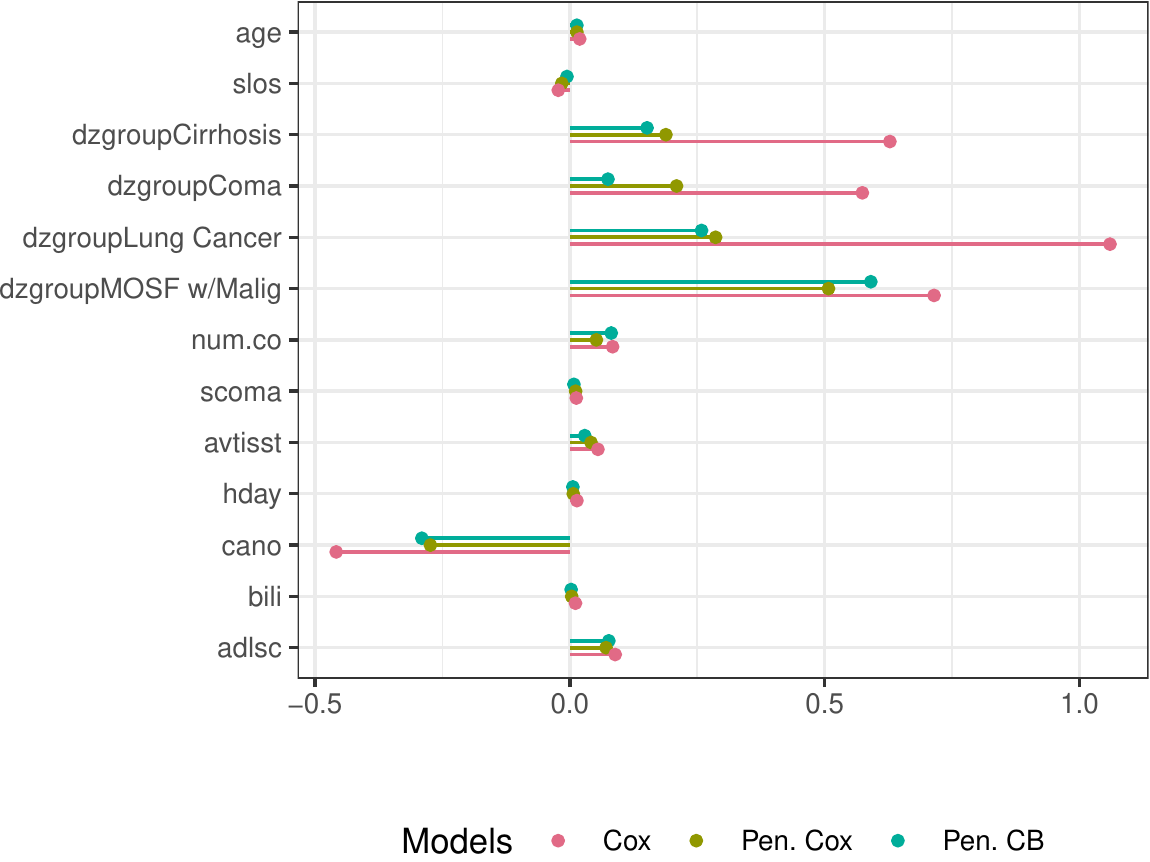}

}

\caption{\label{fig:cs3lolliPlot} Coefficient estimates from the Cox model (Cox), penalized Cox model using the \pkg{glmnet} package (Pen. Cox), and our approach using penalized case-base sampling (Pen. CB). Only the covariates that were selected by both penalized approaches are shown. The shrinkage of the coefficient estimates for Pen. Cox and Pen. CB occurs due to the $\ell_1$ penalty.}\label{fig:coefplots}
\end{figure}
\end{CodeChunk}

We then compare the cumulative incidence estimation over the test set.
The probabilities over time for each observation are averaged, resulting
in the absolute risk curves shown in Figure \ref{fig:cs3FinalBrier} (A).
We can see some minimal differences between the three models, with the
Kaplan-Meier giving the lowest estimates across follow-up-time. Note
that the apparent smoothness of the Cox and penalized Cox curves is due
to the large number of observations in the training set, which is used
to derive the Breslow estimate of the baseline hazard. As described
above, we compare the performance between the models by computing the
Brier scores over time. In Figure \ref{fig:cs3FinalBrier} (B), we can
see that the Brier score is larger for the Kaplan-Meier estimate than
for the other three models. On the other hand, the differences between
these three models are minimal.

\begin{CodeChunk}
\begin{figure}[ht]

{\centering \includegraphics[width=\textwidth,keepaspectratio=true]{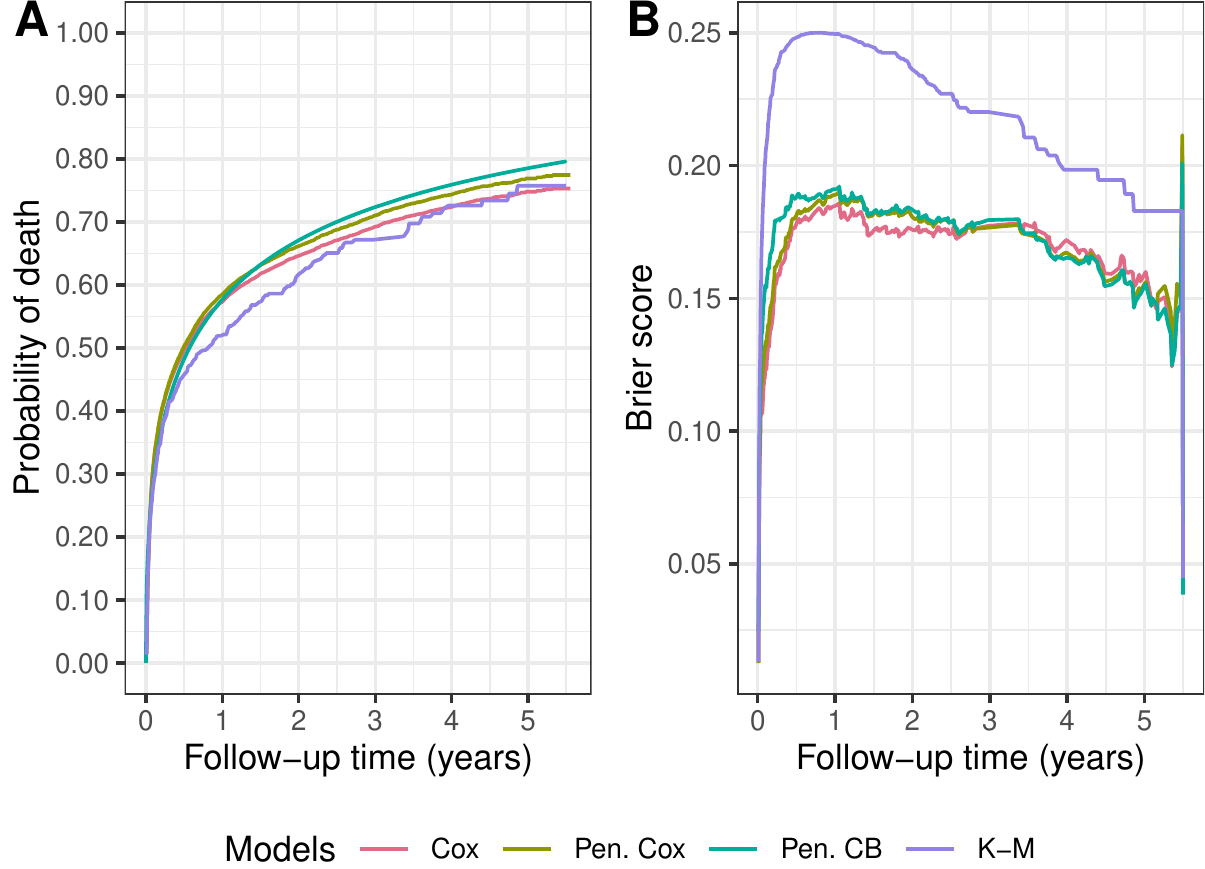}

}

\caption{\label{fig:cs3FinalBrier} Comparison of Cox regression (Cox), penalized Cox regression (Pen. Cox), penalized case-base sampling estimation (Pen. CB), and Kaplan-Meier (K-M). (A)  Probability of death as a function of follow-up time. (B) Brier score as a function of follow-up time, where a lower score corresponds to better performace.}\label{fig:riskregressionBrier}
\end{figure}
\end{CodeChunk}

In this third case study, we showed how case-base sampling can be used
in conjunction with penalized logistic regression to perform variable
selection in survival models.

\hypertarget{case-study-4stanford-heart-transplant-data}{%
\section{Case study 4---Stanford Heart Transplant
Data}\label{case-study-4stanford-heart-transplant-data}}

In the previous case studies, we only considered covariates that were
fixed at baseline. In this next case study, we use the Stanford Heart
Transplant data \citep[\citet{crowley1977covariance}]{clark1971cardiac}
to show how case-base sampling can also be used in the context of
time-dependent exposure. This feature of case-base sampling has been
explored in the literature, in the context of vaccination safety
\citep{saarela2015case}. In this study, the exposure period was defined
as the week following vaccination. Hence, the main covariate of
interest, i.e.~exposure to the vaccine, was changing over time. In this
context, case-base sampling offers an efficient alternative to nested
case-control designs or self-matching.

Recall the setting of Stanford Heart Transplant study: patients were
admitted to the Stanford program after meeting with their physician and
determining that they were unlikely to respond to other forms of
treatment. After enrollment, the program searched for a suitable donor
for the patient, which could take anywhere between a few days to almost
a year. We are interested in the effect of a heart transplant on
survival; therefore, the patient is considered exposed only after the
transplant has occurred.

As above, we can look at the population-time plot for a graphical
summary of the event incidence (see Figure \ref{fig:cs4PopTime}. Here,
we colour the exposed person-time (i.e.~after transplant) in a darker
shade of gray. As we can see, most events occur early during the
follow-up period, and therefore we do not expect the hazard to be
constant.

\begin{CodeChunk}
\begin{figure}[ht]

{\centering \includegraphics[width=\textwidth,keepaspectratio=true]{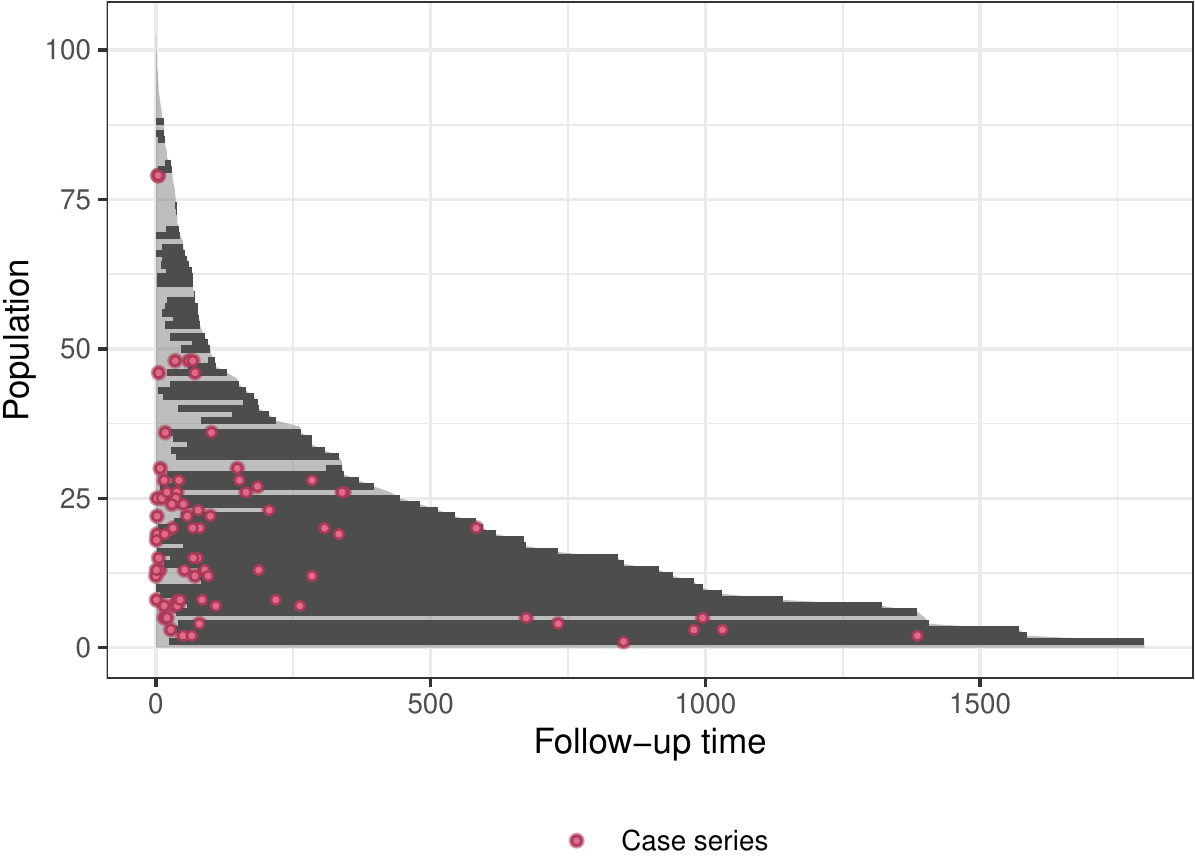}

}

\caption{\label{fig:cs4PopTime} Population-time for the Stanford Transplant study. The darker shade of gray corresponds to the exposed person-time.}\label{fig:stanford-poptime}
\end{figure}
\end{CodeChunk}

Since the exposure is time-dependent, we need to manually define the
exposure variable \emph{after} case-base sampling and \emph{before}
fitting the hazard function. For this reason, we use the
\texttt{sampleCaseBase} function directly.

\begin{CodeChunk}

\begin{CodeInput}
R> cb_data <- sampleCaseBase(jasa,
+   time = "futime",
+   event = "fustat", ratio = 100
+ )
\end{CodeInput}
\end{CodeChunk}

Next, we compute the number of days from acceptance into the program to
transplant, and we use this variable to determine whether each
population-moment is exposed or not.

\begin{CodeChunk}

\begin{CodeInput}
R> # Define exposure variable
R> cb_data <- mutate(cb_data,
+   txtime = time_length(accept.dt %--% tx.date,
+                        unit = "days"
+   ),
+   exposure = case_when(
+     is.na(txtime) ~ 0L, # No transplant
+     txtime > futime ~ 0L,
+     txtime <= futime ~ 1L
+   )
+ )
\end{CodeInput}
\end{CodeChunk}

Finally, we can fit the hazard using various linear predictors.

\begin{CodeChunk}

\begin{CodeInput}
R> # Fit several models
R> fit1 <- fitSmoothHazard(fustat ~ exposure,
+   data = cb_data, time = "futime"
+ )
R> fit2 <- fitSmoothHazard(fustat ~ exposure + futime,
+   data = cb_data, time = "futime"
+ )
R> fit3 <- fitSmoothHazard(fustat ~ exposure * futime,
+   data = cb_data, time = "futime"
+ )
\end{CodeInput}
\end{CodeChunk}

Note that the third model includes an interaction term between exposure
and follow-up time. In other words, this model no longer exhibits
proportional hazards. The evidence of non-proportionality of hazards in
the Stanford Heart Transplant data has been widely discussed
\citep{arjas1988graphical}.

We can then compare the goodness of fit of these three models using the
Akaike Information Criterion (AIC).

\begin{CodeChunk}

\begin{CodeOutput}
#> Model1 Model2 Model3
#>    827    791    790
\end{CodeOutput}
\end{CodeChunk}

As we can see, the best fit is the third model. By visualizing the
hazard functions for both exposed and unexposed individuals, we can more
clearly see how the hazards are no longer proportional. We can easily
obtain a plot of the hazards by using the \texttt{plot.singleEventCB}
method:

\begin{CodeChunk}

\begin{CodeInput}
R> plot(fit3, hazard.params = list(xvar = "futime",
+                                 by = "exposure",
+                                 alpha = 0.05,
+                                 ylab = "Hazard",
+                                 data = cb_data))
\end{CodeInput}
\begin{figure}[ht]

{\centering \includegraphics[width=\textwidth,keepaspectratio=true]{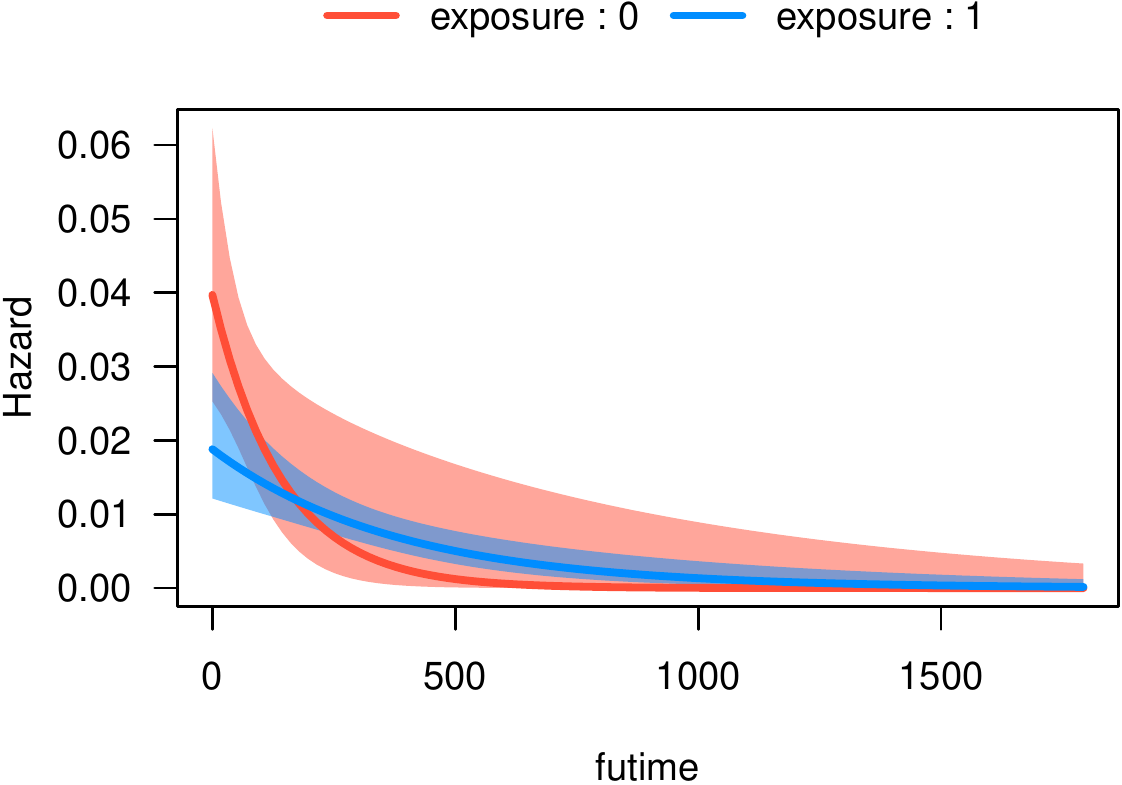}

}

\caption{\label{fig:cs4hazardPlot} Plot of the hazard function for exposed and unexposed individuals.}\label{fig:stanford-hazard}
\end{figure}
\end{CodeChunk}

Looking at Figure \ref{fig:cs4hazardPlot}, the non-proportionality seems
to be more pronounced at the beginning of follow-up than the end.
Finally, in Figure \ref{fig:cs4CIF}, we turn these estimates of the
hazard function into estimates of the cumulative incidence functions.

\begin{CodeChunk}

\begin{CodeInput}
R> # Compute absolute risk curves
R> newdata <- data.frame(exposure = c(0, 1))
R> absrisk <- absoluteRisk(fit3,
+   newdata = newdata,
+   time = seq(0, 425, length.out = 100)
+ )
R>
R> class(absrisk)
\end{CodeInput}

\begin{CodeOutput}
#> [1] "absRiskCB" "matrix"    "array"
\end{CodeOutput}

\begin{CodeInput}
R>
R> plot(absrisk,
+      id.names = c("No Tx", "Tx")) +
+   ylab("Cumulative Incidence") +
+   xlab("Follow-up time (in days)") +
+   paper_gg_theme
\end{CodeInput}
\begin{figure}[ht]

{\centering \includegraphics[width=\textwidth,keepaspectratio=true]{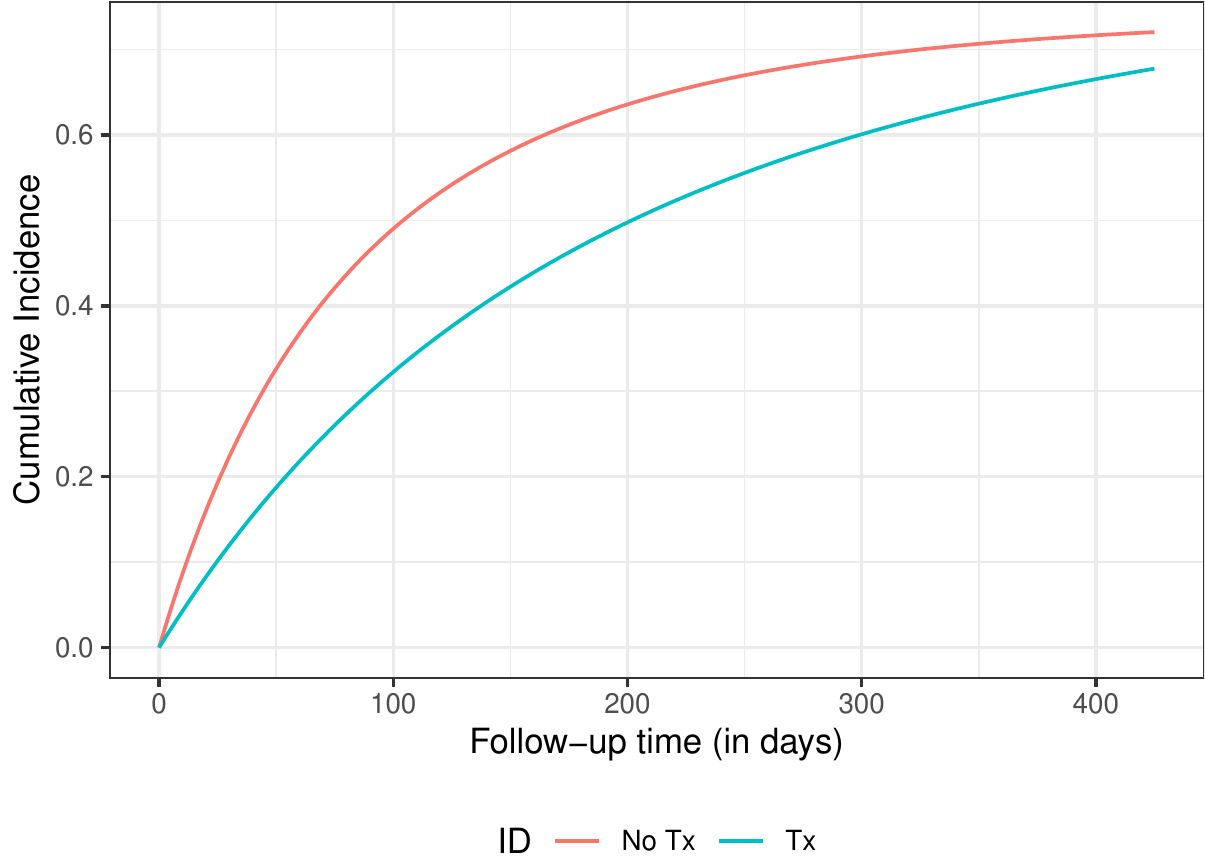}

}

\caption{\label{fig:cs4CIF} Plot of the cumulative incidence function for exposed and unexposed individuals.}\label{fig:stanford-risk}
\end{figure}
\end{CodeChunk}

As we can see in the above case-study, the \pkg{casebase} package can
also be used to model time-varying exposures and non-proportional hazard
functions.

\hypertarget{discussion}{%
\section{Discussion}\label{discussion}}

In this article, we presented the \proglang{R} package \pkg{casebase},
which provides functions for fitting smooth parametric hazards and
estimating CIFs using case-base sampling. Our package also provide
several functions to produce graphical summaries of the data and the
results. We outlined the theoretical underpinnings of the approach, we
provided details about our implementation, and we illustrated the merits
of the approach and the package through four case studies.

As a methodological framework, case-base sampling is very flexible. Some
of this flexibility has been explored before in the literature: for
example, Saarela and Hanley \citeyearpar{saarela2015case} used case-base
sampling to model a time-dependent exposure variable in a vaccine safety
study. As another example, Saarela and Arjas
\citeyearpar{saarela2015non} combined case-base sampling and a Bayesian
non-parametric framework to compute individualized risk assessments for
chronic diseases. In the case studies above, we further explored this
flexibility along two fronts. On the one hand, we showed how splines
could be used as part of the linear predictor to model the effect of
time on the hazard. This strategy yielded estimates of the survival
function that were qualitatively similar to semiparametric estimates
derived from Cox regression; however, case-base sampling led to
estimates of the survival function that \emph{vary smoothly in time} and
are thus easier to interpret. On the other hand, we also displayed the
flexibility of case-base sampling by showing how it could be combined
with penalized logistic regression to perform variable selection.
Furthermore, the second and fourth case studies showed how case-base
sampling can respectively be applied to competing risks and time-varying
exposure settings. Even though we did not illustrate it in this article,
case-base sampling can also be combined with the framework of
\emph{generalized additive models}. This functionality has already been
implemented in the package. Similarly, case-base sampling can be
combined with quasi-likelihood estimation to fit survival models that
can account for the presence of over-dispersion. All of these examples
illustrate how the case-base sampling framework in general, and the
package \pkg{casebase} in particular, allows the user to fit a broad and
flexible family of survival functions.

As presented in Hanley \& Miettinen \citeyearpar{hanley2009fitting},
case-base sampling is comprised of three steps: 1) sampling a case
series and a base series from the study; 2) fit the log-hazard as a
linear function of predictors (including time); and 3) use the fitted
hazard to estimate the CIF. Accordingly, our package provides functions
for each step. Moreover, the simple interface of the
\code{fittingSmoothHazard} function resembles the \code{glm} interface.
This interface should look familiar to new users. Our modular approach
also provides a convenient way to extend our package for new sampling or
fitting strategies.

In the case studies above, we compared the performance of case-base
sampling with that of Cox regression and Fine-Gray models. In terms of
function interface, \pkg{casebase} uses a formula interface that is
closer to that of \code{glm}, in that the event variable is the only
variable appearing on the left-hand side of the formula. By contrast,
both \code{survival::coxph} and \code{timereg::comp.risk} use arrays
that capture both the event type and time. Both approaches to modeling
yield user-friendly code. However, in terms of output, both approaches
differ significantly. Case-base sampling produces smooth hazards and
smooth cumulative incidence curves, whereas Cox regression and Fine-Gray
models produce stepwise CIFs and never explicitly model the hazard
function. Qualitatively, we showed that by using splines in the linear
predictor, all three models yielded similar curves. However, the smooth
nature of the output of \pkg{casebase} provides a more intuitive
interpretation for consumers of these predictions. In Table
\ref{tab:compCBvsCox}, we provide a side-by-side comparison between the
Cox model and case-base sampling.

\begin{table}
\caption{\label{tab:compCBvsCox}Comparison between the Cox model and case-base sampling}
\centering
\begin{tabular}[t]{llp{5cm}}
\toprule
Feature & Cox model & Case-base sampling\\
\midrule
Model type & Semi-parametric & Fully parametric\\
Time & Left hand side of the formula & Right hand side (allows flexible modeling of time)\\
Cumulative incidence & Step function & Smooth-in-time curve\\
Non-proportional hazards & Interaction of covariates with time & Interaction of covariates with time\\
Model testing &  & Use GLM framework \newline (e.g.\ LRT, AIC, BIC)\\
\addlinespace
Competing risks & Difficult & Cause-specific CIFs\\
\bottomrule
\end{tabular}
\end{table}

Our choice of modeling the log-hazard as a linear function of covariates
allows us to develop a simple computational scheme for estimation.
However, as a downside, it does not allow us to model location and scale
parameters separately like the package \pkg{flexsurv}. For example, if
we look at the Weibull distribution as parametrised in
\code{stats::pweibull}, the log-hazard function is given by
\[ \log \lambda(t; \alpha, \beta) = \left[\log(\alpha/\beta) - (\alpha - 1)\log(\beta)\right] + (\alpha - 1)\log t,\]
where \(\alpha,\beta\) are shape and scale parameters, respectively.
Unlike \pkg{casebase}, the approach taken by \pkg{flexsurv} also allows
the user to model the scale parameter as a function of covariates. Of
course, this added flexibility comes at the cost of interpretability: by
modeling the log-hazard directly, the parameter estimates from
\pkg{casebase} can be interpreted as estimates of log-hazard ratios. To
improve the flexibility of \pkg{casebase} at capturing the scale of a
parametric family, we could replace the logistic regression with its
quasi-likelihood counterpart and therefore model over- and
under-dispersion with respect to the logistic likelihood. We defer the
study of the properties and performance of such a model to a future
article.

Future work will look at some of the methodological extensions of
case-base sampling. First, to assess the quality of the model fit, we
would like to study the properties of the residuals (e.g.~Cox-Snell,
martingale). More work needs to be done to understand these residuals in
the context of the partial likelihood underlying case-base sampling. The
resulting diagnostic tools could then be integrated in this package.
Also, we are interested in extending case-base sampling to account for
interval censoring. This type of censoring is very common in
longitudinal studies, and many packages (e.g.~\pkg{SmoothHazard},
\pkg{survival} and \pkg{rstpm2}) provide functions to account for it.
Again, we hope to include any resulting methodology as part of this
package.

In future versions of the package, we also want to increase the
complement of diagnostic and inferential tools that are currently
available. For example, we would like to include the ability to compute
confidence intervals for the cumulative incidence curve. The delta
method or parametric bootstrap are two different strategies we can use
to construct approximate confidence intervals. Furthermore, we would
like to include more functions to compute calibration and discrimination
statistics (e.g.~AUC) for our models. Saarela and Arjas
\citeyearpar{saarela2015non} also describe how to obtain a posterior
distribution for the AUC from their model. Their approach could
potentially be included in \pkg{casebase}. Finally, we want to provide
more flexibility in how the case-base sampling is performed. This could
be achieved by adding a \code{hazard} argument to the function
\code{sampleCaseBase}. In this way, users could specify their own
sampling mechanism. For example, they could provide a hazard that gives
sampling probabilities that are proportional to the cardiovascular
disease event rate given by the Framingham score \citep{saarela2015non}.

In conclusion, we presented the \proglang{R} package \pkg{casebase}
which implements case-base sampling for fitting parametric survival
models and for estimating smooth cumulative incidence functions using
the framework of generalized linear models. We strongly believe that its
flexibility and its foundation on the familiar logistic regression model
will make it appealing to new and established practitioners. The
\pkg{casebase} package is freely available from the Comprehensive
\proglang{R} Archive Network at
\url{https://cran.r-project.org/package=casebase}. Interested users can
visit \url{http://sahirbhatnagar.com/casebase/} for detailed package
documentation and vignettes.

\hypertarget{acknowledgments}{%
\section*{Acknowledgments}\label{acknowledgments}}
\addcontentsline{toc}{section}{Acknowledgments}

We would like to thank Yi Yang for helpful discussions on penalized
regression models and gradient boosting. \mbox{Bhatnagar} gratefully
acknowledges funding via a Discovery Grant from the Natural Sciences and
Engineering Research Council of Canada (NSERC).

\bibliography{references.bib}

\end{document}